\documentclass[aps,prl,preprint,onecolumn,amsfonts,nobibnotes,superscriptaddress,showpacs]{revtex4-1}

\usepackage{dcolumn}% Align table columns on decimal point
\usepackage{amsmath}
\usepackage{amssymb}
\usepackage{graphicx}
\usepackage{bm}
\usepackage{color}
\usepackage{titlesec}
\usepackage{upgreek,xspace}
\DeclareMathOperator{\arcsinh}{arcsinh}

\begin{document}

\title{Clocking the Ultrafast Electron Cooling in\\ Anatase Titanium Dioxide Nanoparticles}

\vspace{2cm}

\author{Edoardo Baldini}
\affiliation{Laboratory of Ultrafast Spectroscopy, ISIC and Lausanne Centre for Ultrafast Science (LACUS), \'Ecole Polytechnique F\'{e}d\'{e}rale de Lausanne, CH-1015 Lausanne, Switzerland}

\author{Tania Palmieri}
\affiliation{Laboratory of Ultrafast Spectroscopy, ISIC and Lausanne Centre for Ultrafast Science (LACUS), \'Ecole Polytechnique F\'{e}d\'{e}rale de Lausanne, CH-1015 Lausanne, Switzerland}

\author{Enrico Pomarico}
\affiliation{Laboratory of Ultrafast Spectroscopy, ISIC and Lausanne Centre for Ultrafast Science (LACUS), \'Ecole Polytechnique F\'{e}d\'{e}rale de Lausanne, CH-1015 Lausanne, Switzerland}

\author{Gerald Aub\"ock}
\affiliation{Laboratory of Ultrafast Spectroscopy, ISIC and Lausanne Centre for Ultrafast Science (LACUS), \'Ecole Polytechnique F\'{e}d\'{e}rale de Lausanne, CH-1015 Lausanne, Switzerland}

\author{Majed Chergui}
\affiliation{Laboratory of Ultrafast Spectroscopy, ISIC and Lausanne Centre for Ultrafast Science (LACUS), \'Ecole Polytechnique F\'{e}d\'{e}rale de Lausanne, CH-1015 Lausanne, Switzerland}

\date{\today}

\begin{abstract}
The recent identification of strongly bound excitons in room temperature anatase TiO$_2$ single crystals and nanoparticles underscores the importance of bulk many-body effects in samples used for applications. Here, for the first time, we unravel the interplay between many-body interactions and correlations in highly-excited anatase TiO$_2$ nanoparticles using ultrafast two-dimensional deep-ultraviolet spectroscopy. With this approach, under non-resonant excitation, we disentangle the optical nonlinearities contributing to the bleach of the lowest direct exciton peak. This allows us to clock the ultrafast timescale of the hot electron thermalization in the conduction band with unprecedented temporal resolution, which we determine to be $<$ 50 fs, due to the strong electron-phonon coupling in the material. Our findings call for the design of alternative resonant excitation schemes in photonics and nanotechnology.
\end{abstract}

%\keywords{}

\maketitle

%Photocatalytic reactions for water splitting, degradation of pollutants and wettability control rely on irradiation of TiO$_2$ substrates with deep-ultraviolet (UV) light, thus generating a two-particle (electron-hole) excitation in the system.

In the last decades, anatase TiO$_2$ has attracted huge interest as one of the most promising materials for a variety of challenging applications, ranging from photocatalysis \cite{ref:fujishima, ref:pelizzetti} and  photovoltaics \cite{ref:oregan} to sensors \cite{bai2014titanium, he2017micro}. Since these technologies involve charge transport, thermalization and localization, they call for studies of the fast electron and hole dynamics, which provide a deep knowledge of the nature of the photogenerated/injected charge carriers and of the energy balance therein. These processes intimately depend on the details of the electronic structure and the presence~ of many body-effects in the material. Since anatase TiO$_2$ is a $d^0$ transition metal oxide, strong electron-electron correlations do not play a substantial role in the electronic structure \cite{cox2010transition}. Hence, this solid can be classified within the simple band insulator scheme, in which the forbidden energy gap arises as a result of band theory and is not a consequence of the strong on-site Coulomb interaction. However, different to conventional band insulators, anatase TiO$_2$ represents a peculiar example in which electron-phonon interaction and electron-hole correlation become relatively strong and influence the optical spectra. On the one hand, the presence of a moderately large electron-phonon coupling in anatase TiO$_2$ has often been invoked to interpret experimental results naturally pointing to the polaronic (self-trapped) picture \cite{ref:tang_urbach, ref:deskins, ref:divalentin, ref:jacimovic, ref:moser, ref:setvin, ref:zhukov_TiO2}. Notable examples include the low temperature green photoluminescence (PL) due to self-trapped excitons \cite{ref:tang_PL, watanabe2000time, watanabe2005time, wakabayashi2005time, harada2007time, cavigli2009volume, cavigli2010carrier} and the room temperature~ electron mobilities whose values are limited by strong scattering with phonons \cite{ref:jacimovic}. On the other hand, many-body correlations have been thought to be negligible in this material and, as such, they remained widely unexplored. Recently, by employing state-of-the-art experimental \cite{ref:baldini_TiO2} and computational techniques \cite{ref:lawler, ref:chiodo, ref:kang, ref:landmann, ref:baldini_TiO2}, the substantial role of electron-hole Coulomb correlations was unravelled in the anatase polymorph of TiO$_2$. Strongly bound direct excitons were found to emerge with a binding energy exceeding 150 meV and a pronounced coupling to the phonon degrees of freedom \cite{ref:baldini_TiO2, toyozawa2003optical}. As a result, these excitons are notably very robust with respect to temperature and sample quality, clearly appearing in single crystals with various degrees of doping, and in defect-rich samples, such as nanoparticles (NPs) with typical radii of 5-15 nm. The latter are representative of the quality of anatase TiO$_2$ used in room temperature~ applications \cite{ref:baldini_TiO2}. 

%These TiO$_2$ NPs, with radii of 5-15 nm, behave similar to the bulk, as the exciton Bohr radius (3.2 nm) rules out quantum confinement effects \cite{ref:baldini_TiO2}. This implies that the electronic properties of anatase TiO$_2$ nanosystems can be analyzed within the framework of the bulk material band structure.

Given the importance of interactions and correlations in anatase TiO$_2$, one expects them to have a pronounced influence on the carrier dynamics. In particular, these phenomena~ are predicted to govern the timescales for intraband carrier cooling via phonon emission \cite{ref:zhukov_TiO2} and they can also influence the optical properties of the material via strong optical nonlinearities. As such, their understanding is of pivotal importance for the design of novel devices and the optimization of existing ones.

So far, the charge carrier dynamics in TiO$_2$ has intensely been studied by ultrafast broadband transient absorption (TA) spectroscopy from the terahertz to the visible \cite{ref:colombo1995, ref:furube1999, ref:yang2001, hendry2004electron, ref:tamaki2006, ref:tamaki2009, ref:yamada2012, ref:matsuzaki2014, zajac}. In these studies, the photoexcited electrons and holes in the system were treated as uncorrelated, an approximation that is valid only when the pump photons are non-resonant with the excitons \cite{ref:baldini_TiO2}. Under these conditions, the material response is dominated by a free-carrier Drude response and by absorption features attributed to localized charges trapped at impurity and/or defect centres \cite{ref:colombo1995, ref:furube1999, ref:yang2001, ref:tamaki2006, ref:tamaki2009, ref:yamada2012}. Although these studies shed light on~ the electron-hole recombination pathways, they were biased towards surface effects, probed only intraband transitions, ignored the material electronic structure, while the identification of impurity/defect bands is still debated \cite{nunzi2016ab, zawadzki2013absorption}. More insightful information was provided by time-resolved PL, which revealed a strongly Stokes-shifted emission in the visible range assigned to self-trapped excitons and/or charges trapped at defects \cite{watanabe2000time, watanabe2005time, wakabayashi2005time, harada2007time, cavigli2009volume, cavigli2010carrier}. However, these processes were found to emerge only at low temperatures and to disappear at room temperature. Furthermore, only long timescales ($>$ 20 ps) were investigated.

To detect possible signatures of interactions and correlations in the dynamical response of anatase TiO$_2$, one needs a TA method that combines a sub-100 fs temporal resolution with both excitation and probing over a broad spectral window in the deep-ultraviolet (UV). It would allow direct access to the region in which spectral weight is removed after above-gap excitation and excitonic features become clearly distinguishable. Insightful information on single-particle and many-body effects is encoded in the bleaching of the exciton state, a nonlinear optical effect that depends on the presence of all the particles (\textit{i.e.} electrons, holes...) in the material \cite{ref:haug, ref:schmitt-rink}.  Such a method has recently become possible as two-dimensional (2D) deep-UV TA spectroscopy \cite{ref:aubock}.

When dealing with excitonic optical nonlinearities in semiconductors, the  most prominent effects are: i) Phase-space filling (PSF), which causes a reduced oscillator strength by decreasing the number of single-particle states contributing to the exciton; ii) long-range Coulomb screening (CS), which broadens the exciton band and shifts it to the blue, as the photoexcited carrier density screens the electron-hole interaction and reduces the exciton binding energy; iii) bandgap renormalization (BGR), which leads to a density-dependent shrinkage of the single-particle states (and consequently of the exciton states), and may reduce the exciton oscillator strength. These processes act simultaneously on the exciton lineshapes, their relative weights being governed by the material parameters and dimensionality \cite{ref:haug, ref:schmitt-rink}. 

Here, using ultrafast 2D deep-UV spectroscopy, we reveal the hierarchy of exciton nonlinearities in anatase TiO$_2$ NPs excited above the band-gap. We find that PSF of the conduction band (CB) by the photoexcited electron density represents the main mechanism upon light generation of uncorrelated electron-hole pairs. This allows us to address the long elusive question of the timescales involved in the intraband hot electron thermalization in anatase TiO$_2$. Specifically, we find an ultrafast electron cooling within 50 fs, which is compatible with phonon-mediated relaxation processes in the strong coupling regime. The practical implication is that excess energy in devices based on anatase TiO$_2$ is ineffective for their functioning and calls for alternative (i.e. resonant) excitation schemes in photonics, photocatalysis and nanotechnology.

The experimental procedures are described in $\S\S$ S1-S2 in the Supporting Information (SI). In a first experiment, we vary the pump between 4.00 eV and 4.60 eV and monitor the TA signal, $\Delta$A,~ over a broad spectral range covering the excitonic resonances (3.50 - 4.60 eV). The time resolution is 150 fs and the photoexcited carrier density is $n$ = 5.7 $\times$ 10$^{19}$ cm$^{-3}$. This high-excitation density regime is chosen to avoid polaronic effects within the bands and promote the formation of a mobile electron-hole liquid \cite{ref:moser}. Figure 1(a) shows ultrafast 2D deep-UV pump-probe colour-coded maps at different time delays (1 ps, 100 ps and 500 ps). The straight line in each map indicates the region in which pump and probe have the same photon energy, and it is affected by artefacts due to the scattering of the pump beam into the spectrometer. All transients are characterized by a negative signal over the entire probe spectral range, in which two long-lived features are clearly distinguished at probe photon energies of 3.88 eV and 4.35 eV. While the former is present at all pump photon energies, the latter becomes more prominent when the pump is tuned above 4.10 eV. Cuts of the 1 ps map at specific probe photon energies are shown in Fig. 1(b), normalized to the maximum amplitude of the 3.88 eV feature. They are compared to the inverted steady-state absorption spectrum ($-A$) of the same colloidal solution of anatase TiO$_2$ NPs (black circles). In contrast to the latter, both features appearing in the $\Delta$A spectra have a well-defined Lorentzian lineshape with a width of $\sim$300 meV. These bands are hidden in the steady-state absorption spectrum \cite{ref:serpone} due to the strong scattering of light, but become apparent in the TA spectrum due to cancellation of the probe scattered light in the difference spectrum. Moreover, the peak of the low-energy Lorentzian band is independent of the pump photon energy, while clear information on the high-energy feature cannot be retrieved due to the influence of the pump beam scattering. We recently identified the  features at 3.88 eV and $\sim$ 4.35 eV as bound excitons along the a- and c-axis of single crystals of anatase TiO$_2$ \cite{ref:baldini_TiO2}. They show up in Fig. 1 due to the random orientation of NPs in solution, underscoring the analogy of the spectroscopic features of NPs and bulk single crystals.

From Fig. 1 we can already draw an important conclusion on the nonequilibrium dynamics of anatase TiO$_2$ NPs, which was neglected in previous studies. Depending on the choice of the pump photon energy, one can create uncorrelated electron-hole pairs (non-resonant excitation) or excitons (resonant excitation). In the former case, the pump photon energy determines the regions of the Brillouin zone where the nonequilibrium electron-hole density is created. This consideration is of importance in the case of an indirect bandgap insulator such as anatase TiO$_2$, since the pump photon energy can also be tuned to phonon-assisted interband transitions (despite their lower cross section). On the contrary, resonant excitation leads to the creation of bound exciton species in the system, which complicates the interpretation of the ultrafast dynamics due to the presence of many-body exciton-exciton interactions. This case will be addressed in future work, while here we focus on the non-resonant excitation and clarify the impact of uncorrelated electron-hole pairs on the bound exciton spectral feature. To this purpose, we select a data set from Fig. 1, namely the one at the pump photon energy of 4.05 eV, since it lies above the a-axis~ and below the c-axis exciton peak \cite{ref:baldini_TiO2} and is expected to minimize their contribution. This, in turn, allows us to link more accurately the resulting ultrafast dynamics to the details of the electronic structure. The data are displayed in Fig. 2(a) as a colour-coded map of $\Delta$A as a function of probe photon energy and time delay between pump and probe. As expected, the spectral response features the long-lived a-axis exciton band at 3.88 eV, followed by the weaker shoulder of the c-axis exciton around 4.35 eV. The transient spectra at different time delays of Fig. 2(b) show that the 3.88 eV band does not change with time (Fig. S2). Figure 2(c) displays the kinetic traces at three specific probe photon energies: below (3.77 eV), at (3.90 eV) and above (4.10 eV) the a-axis exciton peak. All temporal traces exhibit a resolution-limited rise of 150 fs, followed by a long-lived decay persisting beyond 1 ns. In addition, a low-frequency oscillation modulates the whole spectrum during the first 5 ps, which is fully damped within one oscillation. This feature is due to coherent acoustic phonons confined within the spherical TiO$_2$ NPs, and it will be discussed elsewhere. Here, we retrieve the significant parameters of the incoherent response, by performing a global fit of sixteen selected temporal traces in the 3.60-4.40 eV probe range of the $\Delta$A map. A satisfactory global fit~ up to 1 ns (solid line in Fig. 2(c)) is optimal using a multiexponential function with four time constants ($\tau$) convoluted with a Gaussian accounting for the instrument response function (IRF) of $\sim$ 150 fs: $\tau_1$ = 1.60 $\pm$ 0.12 ps, $\tau_2$ = 10 $\pm$ 0.40 ps, $\tau_3$ = 50 $\pm$ 1.70 ps, $\tau_4$ = 423 $\pm$ 14.70 ps. The global fit~ enables us to disentangle the spectral dependence of the decay processes contributing to the recovery of the exciton bleach. To this aim, in Fig. 2(d), we plot the pre-exponential factors associated with a given time constant as a function of the probe photon energy. The spectral dependence of the $\tau_1$ and $\tau_2$ components show a negative amplitude broader than the exciton lineshapes, while the contributions of the $\tau_3$ and $\tau_4$ components mainly reproduce the contours of both exciton bands, suggestive of a different nature for the two sets of components. All decay processes are strictly related to the electron-hole recombination mechanisms, which reduce the density of delocalized carriers via radiative or non-radiative relaxation channels. 

In order to identify the phenomena behind the different time constants, one first needs to assess the relative weights of the optical nonlinearities contributing to the transient signal. Central for this is to clarify the electron-hole populations giving rise to the exciton collective state. Based on the band structure of anatase TiO$_2$ \cite{ref:chiodo}, the single-particle states contributing to the a-axis exciton lie along the $\Gamma$-Z direction of the Brillouin zone (Fig. 3). Due to the symmetry of the $p$-$d$ wavefunctions, however, the states exactly at the $\Gamma$ point are symmetry forbidden and do not contribute to the exciton state \cite{ref:chiodo}. Once a nonequilibrium distribution of uncorrelated electron-hole pairs is created, the electrons are expected to thermalize to the bottom of the CB at $\Gamma$ and the holes to the top of the valence band (VB) close to X. Since the deep-UV probe is sensitive to the joint density of states in the material, an induced transparency ($\Delta$A $<$ 0) can arise from a density of electrons (holes) accumulated in the CB (VB) alone. In anatase TiO$_2$, the PSF contribution to the exciton bleaching is expected to arise exclusively due an electron population close to the bottom of the CB. The hole contribution is absent, due to their ultimate relaxation to the top of the VB close at the X point. Consistent with this scenario, the exciton bleach is found to persist even when the pump photon energy is tuned to 3.54 eV, thus promoting indirect (phonon-assisted) interband transitions. On the contrary, upon 3.10 eV excitation, no signal is detected even up to incident fluences of 34 mJ/cm$^2$.

The above considerations imply that PSF, long-range CS and BGR may all contribute to the observed optical nonlinearities. This requires addressing the effects related to the enhancement of the electronic screening upon pump photoexcitation. To this aim, we study how the exciton spectrum is renormalized as a function of the pump fluence at 4.05 eV, narrowing our detection range to cover only the spectral region of the a-axis exciton. This allows a higher stability of our setup, maintaining the same experimental conditions over the time required for the fluence dependence study. Within the explored range, the maximum intensity of the signal~ scales linearly with the absorbed fluence (Fig. S3), which excludes multiphoton absorption processes from the pump beam. The normalized spectra are shown in Fig. 4(a,b) for delay times of 400 fs and 10 ps, respectively. We observe that the exciton lineshape slightly broadens with increasing carrier density. We assign this behaviour to the presence of long-range CS. Indeed, under our experimental conditions, we can exclude that the broadening originates from other exciton decay channels opened by the photoexcitation process. Radiative or non-radiative exciton decay processes are found in direct bandgap semiconductors in the presence of low-dimensionality and reduced dielectric screening \cite{fuchs1993auger, klimov2000quantization, wang2006auger, sun2014observation}, since non-resonant photoexcitation can spontaneously evolve into exciton formation via the single-particle states at the band edges. In contrast, in indirect bandgap semiconductors, photoexcited uncorrelated electrons and holes quickly relax towards the bottom of their respective bands and no direct excitons are energetically favoured \cite{schweizer}. In principle, indirect excitons would be allowed to form through the mediation of phonon modes. However, in the case of anatase TiO$_2$ nanosystems, their existence at room temperature has been disproven by extensive measurements \cite{zhang2000photoluminescence, harada2007time, knorr2008trap, mercado2012location} and by the results we present in the following. As such, many-body processes such as exciton-exciton annihilation can be excluded. Consistent with the idea that long-range CS from free carriers is the optical nonlinearity governing the exciton broadening, in a recent study we found that the excitonic band of anatase TiO$_2$ is broadened upon charge injection from an external dye adsorbed on the surface of NPs (i.e. in the absence of holes in the VB of TiO$_2$) \cite{baldini_dye}. Long-range CS is also expected to lead to a blueshift of the exciton peak. However, within our experimental accuracy, here we observe no shifts in the peak energy position, nor derivative-like shapes. This apparent insensitivity of the exciton peak energy to the photoinduced carrier density even in the presence of CS-induced broadening has been rationalized in the literature as an exact cancellation of the effects of CS and BGR on the excitonic resonances at all densities \cite{ambigapathy1997coulomb, reynolds2000combined, sarma2000many, stopa2001band} . In general, the quantitative details of this compensation depend on both material and dimensionality. From these measurements, we can conclude that the carrier density which is required to induce dramatic changes of the exciton peak energy is higher than those produced by our photoexcitation. As a consequence, at sufficiently low electron-hole pair densities, when the continuum is still far from the resonances, only the loss of oscillator strength due to PSF and a slight broadening due to CS are apparent. %These results further confirm the strongly bound exciton scenario deduced for the a-axis exciton in Ref. 21, which also demonstrated the ineffectiveness of BGR even for extremely high carrier densities ($n \sim$ 10$^{20}$ cm$^{-3}$) \cite{ref:baldini_TiO2}. Thus, we conclude that PSF and CS are the dominant effects contributing to the excitonic optical nonlinearities at our carrier densities. 

Having established the dominant role played by CB electrons in blocking the excitonic transitions allows one to retrieve valuable information on the electron recombination dynamics. Indeed, the $\Delta$A signal at the excitonic resonance can be used as a measure of the photoexcited electron concentration changing with time. Representative temporal traces at a probe photon energy of 3.88 eV and for different excitation densities are shown in Fig. 4(c) and normalized with respect to their maximum. We observe that below 40 ps the bleach recovery accelerates with fluence, which is indicative of higher-order recombination processes for the charge carriers, such as bimolecular and Auger recombination. Indeed, the recombination dynamics in semiconductors and insulators proceeds via single-carrier nonradiative processes (trapping at impurity states), two-body radiative (bimolecular) mechanisms and nonradiative trap-Auger recombination processes and three-body band-to-band Auger processes~\cite{ref:landsberg}. Since anatase TiO$_2$ is an indirect bandgap insulator, band-to-band radiative recombination is known to be extremely inefficient. Thus, the only radiative recombination pathways that can take place at room temperature~ are those involving a delocalized carrier in a band and a localized carrier trapped at a defect state. To verify the efficiency of the radiative recombination processes, we measured the PL obtained in colloidal anatase TiO$_2$ NPs by femtosecond fluorescence up-conversion. As it is well established that the spectral content of the PL does not depend on the pump photon energy for above-gap excitation, we illuminate the NPs with a pump pulse centered around 4.66 eV. The broad PL spectrum at a time delay of 1 ps is shown in Fig. \ref{fig:PL}(a). The PL appears only in the visible regime and is completely absent in the spectral region $>$ 3.00 eV. It retains the form of a broad band centered around 2.24 eV, which is characterized by an extremely weak intensity. Figure \ref{fig:PL}(b) displays the temporal traces at 1.91 eV, 2.18 eV and 2.58 eV. The PL signal rises within our experimental temporal resolution (200 fs) and decays bi-exponentially with time constants of 2 ps and 30-40 ps. Since the PL band is centered around 2.24 eV, it can be readily assigned to an extrinsic radiative recombination channel that involves carriers trapped at defect states, ruling out the involvement of any self-trapped exciton recombination process (see SI). By evaluating the PL quantum yield for time delays below 500 ps, we find it to be of the order of $\sim$ 1.3 $\times$ $10^{-6}$, i.e. extremely weak. As a consequence, it is straightforward to assume that the recombination dynamics at early time delays in anatase TiO$_2$ is entirely governed by Auger mechanisms, as already inferred from TA spectroscopy studies in the infrared and the visible ranges for highly-excited anatase TiO$_2$ NPs and thin films~\cite{ref:colombo1995, ref:matsuzaki2014}. It is also consistent with the spectral dependence of the $\tau_1$ and $\tau_2$ relaxation components (Fig. 2(d)), whose broad shape hints to energetically redistributed carriers over a wider phase space during the Auger processes. In contrast, the longer $\tau_3$ and $\tau_4$ relaxation components can be assigned to electron trapping processes at defect states, which lead to bleach recovery by emptying the phase space involved in the exciton state.

More interestingly, having established PSF as the main mechanism behind exciton bleaching allows us to address the long elusive issue of the timescale of electron cooling to the CB minimum in anatase TiO$_2$. A recent single-wavelength transient reflectivity experiment on rutile TiO$_2$ single crystals tracked the influence of carrier cooling onto the phase of the coherent $A_g$ phonon mode at 74.4 meV, offering valuable information on the electron-phonon coupling dynamics \cite{ref:bothschafter}. Nevertheless, this approach is blind to the details of the electronic structure and relies on the analysis of coherent optical phonons, whose observation is not always straightforward.

Here, we circumvent these limitations by demonstrating the effectiveness of exciton bleaching as a probe of the intraband electron thermalization. To this aim, we focus on the rise of the exciton bleach signal by reducing the IRF of our setup to 80 fs (see Methods). In Fig. 6(a), we compare the signal at 3.88 eV and an excitation density $n \sim$ 2.1 $\times$ 10$^{19}$ cm$^{-3}$ (violet curve) with that of the pure solvent (pink curve). In both traces, artefacts due to residual cross-phase modulation~ (CPM) after t = 0 can be identified. Moreover, while the signal from the solvent dies immediately after the CPM, the one from anatase TiO$_2$ persists over time and comprises a relaxation component that starts around 200 fs. However, to isolate the anatase TiO$_2$ signal, the two traces cannot be directly subtracted due to slightly different experimental conditions in the measurements. Therefore, we adopt a common procedure in ultrafast spectroscopy, in which the IRF is assumed to coincide with the duration of the CPM signal and is represented by a Gaussian function (with a full-width at half-maximum of 80 fs) folding in the CPM modulations (blue curve in Fig. 6(a)). The isolated TiO$_2$ signal is shown as red dots superimposed to the original trace. To provide an upper limit to the rise time of the bleach signal, in Fig. 6(b) the isolated TiO$_2$ signal (red dots) is compared to three computed time traces convoluted with the Gaussian IRF of 80 fs. They consist of different rise times and an exponential recovery time of 200 fs. Here, the time t = 0 is consistently defined with respect to the subtracted CPM. It can be seen that the 20 fs rise time interpolates nicely between the t $<$ 0 points and the t $>$ 200 fs ones. In any case, the upper limit cannot exceed 50 fs. Thus, the rise of the exciton bleach is $<$ 50 fs, which is the timescale for the intraband electron cooling.

We now discuss the origin of this ultrafast electron cooling. Upon photoexcitation, momentum conservation results in the partition of the excess energy provided by the pump pulse into kinetic energy of the electrons and holes according to the reciprocal ratio of their effective mass $m_e^*$/$m_h^*$, with the carrier having lower effective mass receiving more excess energy

\begin{equation}
E\mathrm{^e_{exc}}(\hbar\omega) = \frac{\hbar^2 k^2}{2m_e^*} = \frac{m_h^*}{m_e^* + m_h^*}\left( \hbar\omega - E\mathrm{_{gap}}\right) 
\end{equation}

\begin{equation}
E\mathrm{^h_{exc}}(\hbar\omega) = \frac{\hbar^2 k^2}{2m_h^*} = \frac{m_e^*}{m_e^* + m_h^*}\left( \hbar\omega - E_\mathrm{_{gap}}\right) 
\end{equation}
\newline
\noindent In our experiment, $\hbar\omega$ = 4.05 eV and $E\mathrm{_{gap}}$ = 3.20 eV. Rigorously, the electron and hole effective masses vary as a function of momentum $k$. These masses are well-defined concepts only within the parabolic approximation for the band structure. From the latter, we expect our pump photons to promote uncorrelated electron-hole pairs in the vicinity of the $\Gamma$ point along the $\Gamma$-X direction (purple arrow in Fig. 3). Due to the high carrier densities involved in our experiment, we use the value of the electron effective mass reported by band theory $m_e^*$ = 0.42$m_e$ (see \S S5 for the choice of $m_e^*$) \cite{hitosugi2008electronic}. On the other hand, to provide an estimate of $m_h^*$ from the details of the measured electronic structure, we rely on recent ARPES data for the top of the VB along the $\Gamma$-X direction in the case of an excess carrier density $n$ $\sim$ 10$^{19}$ cm$^{-3}$ \cite{ref:baldini_TiO2}. Performing a parabolic fit yields $m_h^*$ = (2.78 $\pm$ 0.5)$m_e$, which also accounts for the polaronic contribution to the mass renormalization. The values of $m_e^*$ and $m_h^*$ lead to the excess energy partition $E^e_{exc}$ $\sim$ 0.74 eV and $E^h_{exc}$ $\sim$ 0.11 eV.

Once the carriers have received this excess energy in their respective bands, they start interacting with the phonon modes of the lattice. When the timescale obtained for the electron cooling in anatase TiO$_2$ is compared with the slower ($>$ 250 fs) electron relaxation timescales retrieved in nonpolar indirect gap semiconductors, some conclusions can be drawn concerning the role of electron-phonon scattering in different materials. In nonpolar semiconductors (such as Ge and Si) phonon scattering occurs through the optical phonon deformation potential interaction, while in partially-ionic polar systems (such as anatase TiO$_2$) the most effective coupling mechanism is represented by the electron-longitudinal optical (LO) phonon scattering described by the Fr\"ohlich interaction \cite{frohlich1950xx}. In TiO$_2$, the optical phonons that are most strongly coupled to the electronic degrees of freedom are those belonging to the branches of the $E_u$ and $A_{2u}$ modes \cite{ref:deskins, ref:moser, baldini_rutile}. Due to this coupling, in the low carrier density limit, the CB electrons form well-defined large polaron quasiparticles, characterized by a dimensionless polaron coupling constant $\alpha$ $\sim$ 2 \cite{ref:moser}. This constant determines the ratio between the polaron self-energy and the LO phonon energy, thus defining an intermediate-to-strong electron-phonon coupling regime for the electrons in anatase TiO$_2$. At high density, the polaronic quasiparticle are observed to collapse into an electron liquid coupled to the phonon modes. Within the single-particle limit probed by ARPES, the crossover takes place around $n$ = 10$^{19}$ cm$^{-3}$ \cite{ref:moser}. In the two-particle limit of optical absorption, this threshold may be even lower, leading to the emergence of a well-defined Drude response for the photocarriers \cite{ref:matsuzaki2014} instead of the characteristic absorption features of large polarons \cite{devresee}. Even in this diffusive regime, the electron-phonon coupling can lead to an efficient and fast transfer of electron excess energies to the phonon bath within a few tens of fs.

To theoretically interpret the relaxation times observed in our experiment, we first rely on a simple estimate of the scattering rate as expected from first order perturbation theory. In the case of electrons in a parabolic CB, the scattering rate can be calculated via Fermi's golden rule as \cite{ridley2013quantum}
\begin{equation}
\Gamma = \Gamma_0 \Big(\frac{1}{\epsilon_\infty}-\frac{1}{\epsilon_s}\Big)\arcsinh\Big(\sqrt{\frac{E^e_{exc}}{\hbar\omega_{LO}} -1}\Big)(1+2n_{BE})
\end{equation}

\noindent where $\epsilon_{\infty}$ and $\epsilon_{s}$ are the dielectric constants at energies well above and below the phonon energy $\omega_{LO}$, respectively, and $n\mathrm{_{BE}}$ is the Bose-Einstein statistical factor. $\Gamma_0$ is a nearly temperature-independent prefactor that reads
\begin{equation}
\Gamma_0 = \sqrt{\frac{m_e^*}{2E^e_{exc}}}\frac{e^2 \omega_{LO}}{2\pi\hbar\epsilon_0}
\end{equation}

Substituting the relevant parameters for anatase TiO$_2$ \cite{gonzalez1997infrared} yields $\sim$ 4 fs for the electron cooling time, which could be compatible with our experimental observation. However, we point out that Eq. (3) is valid under the conditions that a single phonon mode is interacting with the electrons and that $\alpha$ $<$ 1. Since $\alpha$ represents a measure for the relative importance of higher-order processes, Eq. (3) may not provide an accurate estimate of the actual electron cooling rate. A more reliable estimate relies on density-functional perturbation theory accounting for the phonon density of states of the material, as computed in Ref. \cite{ref:zhukov_TiO2}. According to this \textit{ab initio} approach, under our experimental conditions we expect the timescale for emission of a single phonon by an excited electron to be $\sim$2 fs and the total thermalization time to the CB edge $\sim$40 fs, in very good agreement with the present upper limit of 50 fs (Fig. 3). On the other hand, the holes can undergo rapid intervalley scattering to the X and Z points via emission of high-wavevector phonons.

Within the semiclassical limit, for excess energies less than the minimum LO phonon energy ($\sim$ 45 meV in TiO$_2$ \cite{gonzalez1997infrared}), severe constraints in the phonon phase space are expected to result in a dramatic decrease of the intraband cooling rate \cite{ref:zhukov_TiO2}. However, for polaronic materials governed by the Fr\"ohlich interaction, purely quantum kinetic relaxation channels can open also in the case of small excess energies, leading to an efficient redistribution of the electronic energy into the strongly coupled phonons \cite{betz1999ultrafast, betz2001subthreshold}. This effect can be viewed as the buildup of the polaronic dressing in the low-density limit. Future quantum kinetic calculations involving the solution of the Dyson equation will shed light on these phenomena. At this stage, irrespective of the mechanism at play, the net result is that no effective energy will be stored in the electronic degrees of freedom for sufficiently long timescales.

These arguments lead us to conclude that, in anatase TiO$_2$, fast relaxation processes impose serious limitations to the practical use of the full photon energy in applications. This scenario is radically different from the physics of semiconductor quantum dots, in which the spacing between the discrete electronic levels is large enough to prevent fast phonon-mediated carrier cooling processes \cite{bockelmann1990phonon}. For photocalaysis, fast thermalization is a favourable process, since the lower energy states are important. This issue is also of importance in photovoltaics since a large driving force is often sought after for efficient injection, especially when the coupling between the sensitizer to the TiO$_2$ substrate is weak. However, the present study shows that the large excess energy is lost to phonons. Therefore, alternative excitation schemes, such as resonant excitation of the strongly bound exciton species, are preferable in which the energy can be temporarily harvested. Another ideal situation is illustrated by the simulations by de Angelis \textit{et al.} \cite{de2011absorption}, showing a strong hybridization of the wavefunctions of the sensitizer and the substrate. Furthermore, as we showed that the a-axis excitons are 2D and robust against temperature and defects, they are expected to move freely on the (001) plane \cite{ref:chiodo, ref:baldini_TiO2}. As such, they may offer an efficient source to mediate the flow of energy at the nanoscale in engineered devices based on anatase TiO$_2$.

In summary, we have provided direct evidence of excitonic optical nonlinearities in anatase TiO$_2$ nanoparticles at room temperature. Besides their fundamental importance, these excitons hold huge promise for future technological developments, in particular in the emerging field of excitonics. We revealed selective information on the intraband and interband electron dynamics by tracking the renormalization of the excitonic features over time. We showed that in nanosized anatase TiO$_2$ the strong electron-phonon coupling of the material results in very fast cooling times for electrons possessing a substantial ($>$ 0.2 eV) excess energy. Under these conditions, any optical or acoustic phonon can be excited in the course of the electron thermalization, thus contributing to a rapid loss of the electronic excess energy, making it ineffective in devices based on anatase TiO$_2$. More remarkably, our results highlight the importance of relating the carrier dynamics of anatase TiO$_2$ NPs to the details of the electronic structure, an aspect that was overlooked in all previous experimental studies. Finally, by unraveling the origin of the bound exciton bleaching in NPs, our findings pave the way to its use as a sensor of carrier dynamics in a variety of nanosystems based on anatase TiO$_2$. Notable examples include modified forms of this material, such as the ones obtained upon chemical doping with magnetic ions or upon crystal structure control.

\begin{acknowledgments}
We thank Prof. Angel Rubio for the critical reading of our paper and Prof. Anders Hagfeldt for valuable discussions. We acknowledge the Swiss NSF for support via the NCCR:MUST and the contracts No. 206021-157773, 20020-153660 and 407040-154056 (PNR 70), the European Research Council Advanced Grants H2020 ERCEA 695197 DYNAMOX.
\end{acknowledgments}

\newpage
\clearpage

\begin{figure}[t]
	\centering
	\includegraphics[width=\linewidth]{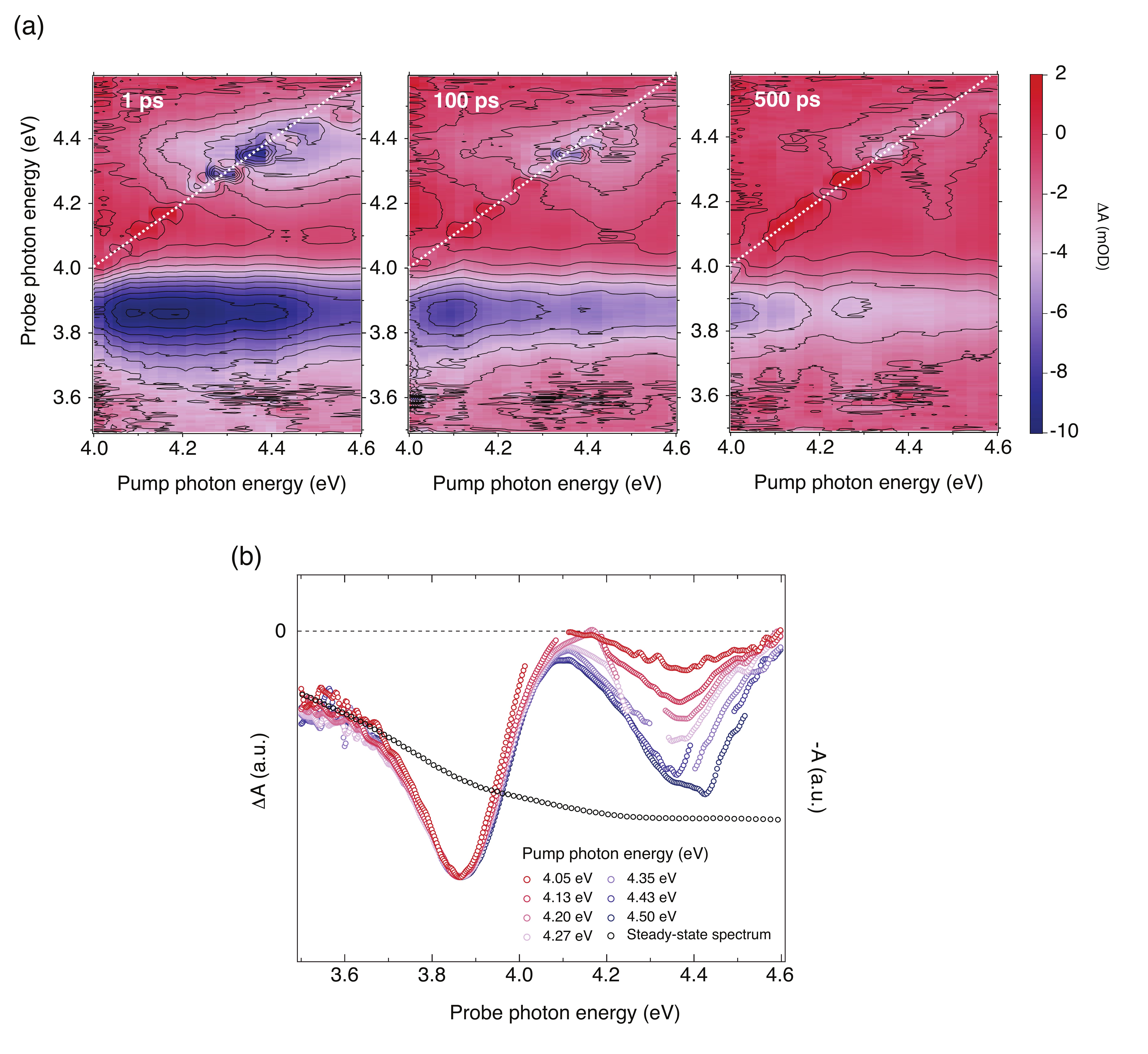}
	\caption{\textbf{(a)} Colour-coded maps of $\Delta$A measured via ultrafast 2D deep-UV spectroscopy on a colloidal solution of anatase TiO$_2$ NPs at room temperature~ as a function of pump- and probe photon energy. The spectral response is displayed at three different time delays between pump and probe: 1 ps, 100 ps, 500 ps. The time resolution is estimated to be 150 fs and the photoexcited carrier density is $n$ = 5.7 $\times$ 10$^{19}$ cm$^{-3}$. \textbf{(b)} Normalized $\Delta$A spectra at a fixed time-delay of 1 ps and for different pump photon energies (indicated in the figure). Each trace is normalized with respect to the minimum of the main feature at 3.88 eV. For comparison, the black trace shows the inverted ($-A$) steady-state absorption spectrum.}
\end{figure}
\newpage

\begin{figure}[t]
	\centering
	\includegraphics[width=\linewidth]{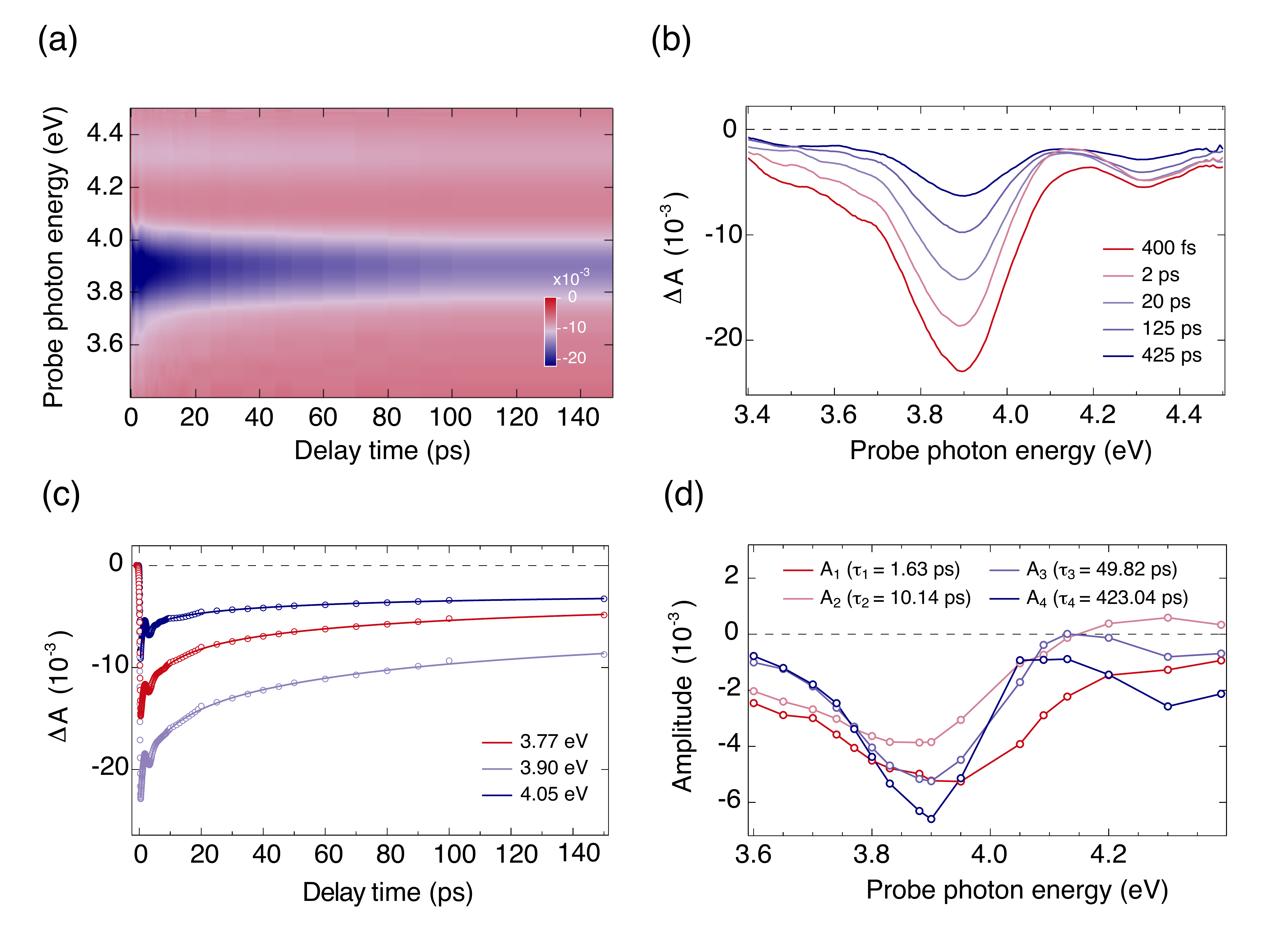}
	\caption{\textbf{(a)} Colour-coded map of $\Delta$A measured on a colloidal solution of anatase TiO$_2$  NPs as a function of probe photon energy and time delay between pump and probe. The time resolution is estimated to be 150 fs, the pump photon energy is set at 4.05 eV and the photoexcited carrier density is $n$ = 5.7 $\times$ 10$^{19}$ cm$^{-3}$. \textbf{(b)} $\Delta$A spectra as a function of probe photon energy at representative delay times between pump and probe. \textbf{(c)} Experimental temporal traces of $\Delta$A for different probe photon energies (dotted lines) and results of the global fit analysis (solid lines). \textbf{(d)} Contribution to the $\Delta$A response of the four relaxation components obtained from the global fit analysis.}
\end{figure}
\newpage

\begin{figure}[t]
	\centering
	\includegraphics[width=0.5\linewidth]{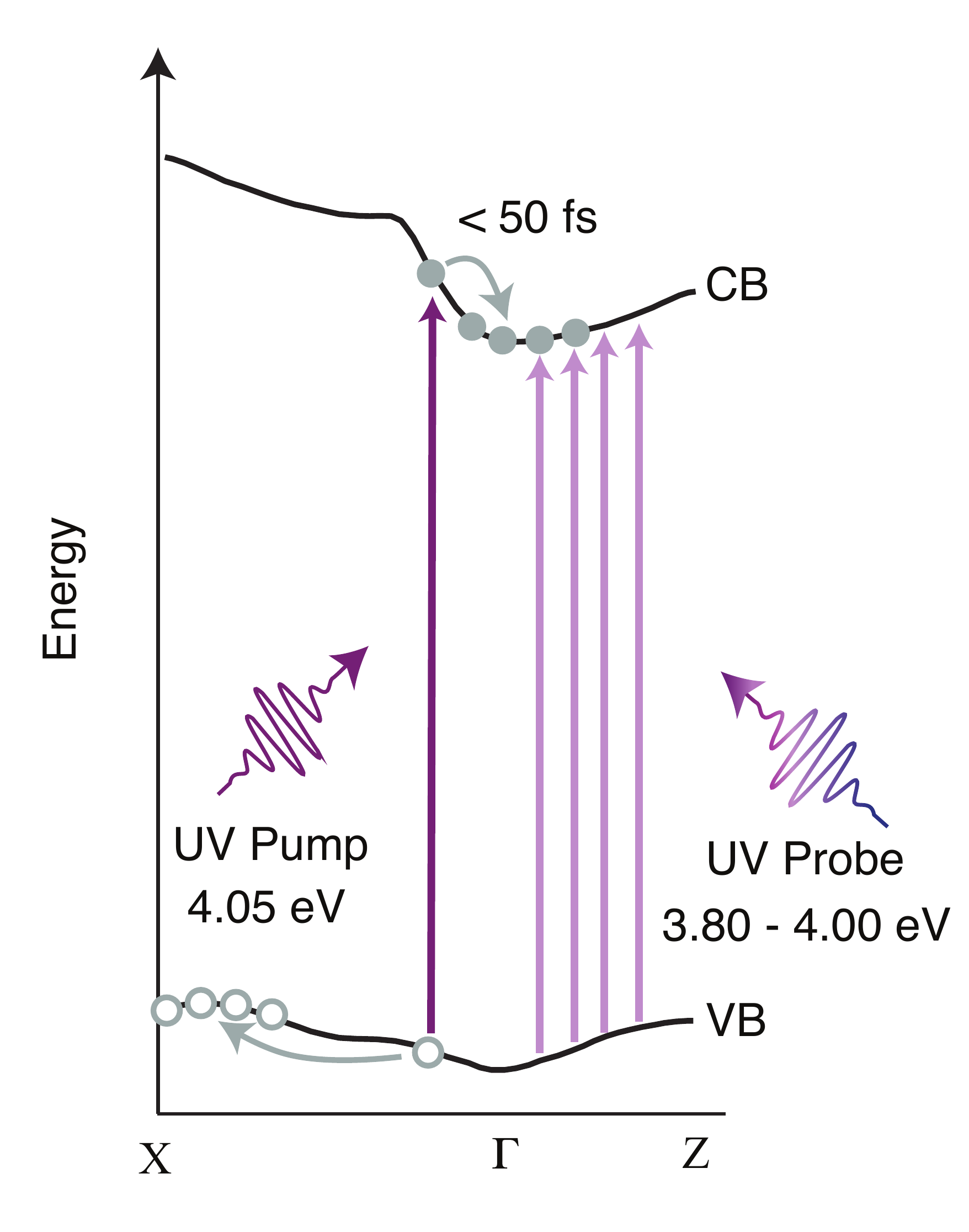}
	\caption{Schematic representation of the deep-UV based detection of the ultrafast carrier dynamics in anatase TiO$_2$ NPs. The pump photon at 4.05 eV (purple arrow) excites electron-hole pairs through direct transitions. The broadband UV pulse (violet arrows) probes the exciton feature at 3.88 eV. The direct transitions contributing to this collective state lie along the $\Gamma$-Z region of the Brillouin zone. The band structure of anatase TiO$_2$ has been adapted from Ref. \cite{ref:chiodo}.}
\end{figure}
\newpage

\begin{figure}[t]
	\centering
	\includegraphics[width=\linewidth]{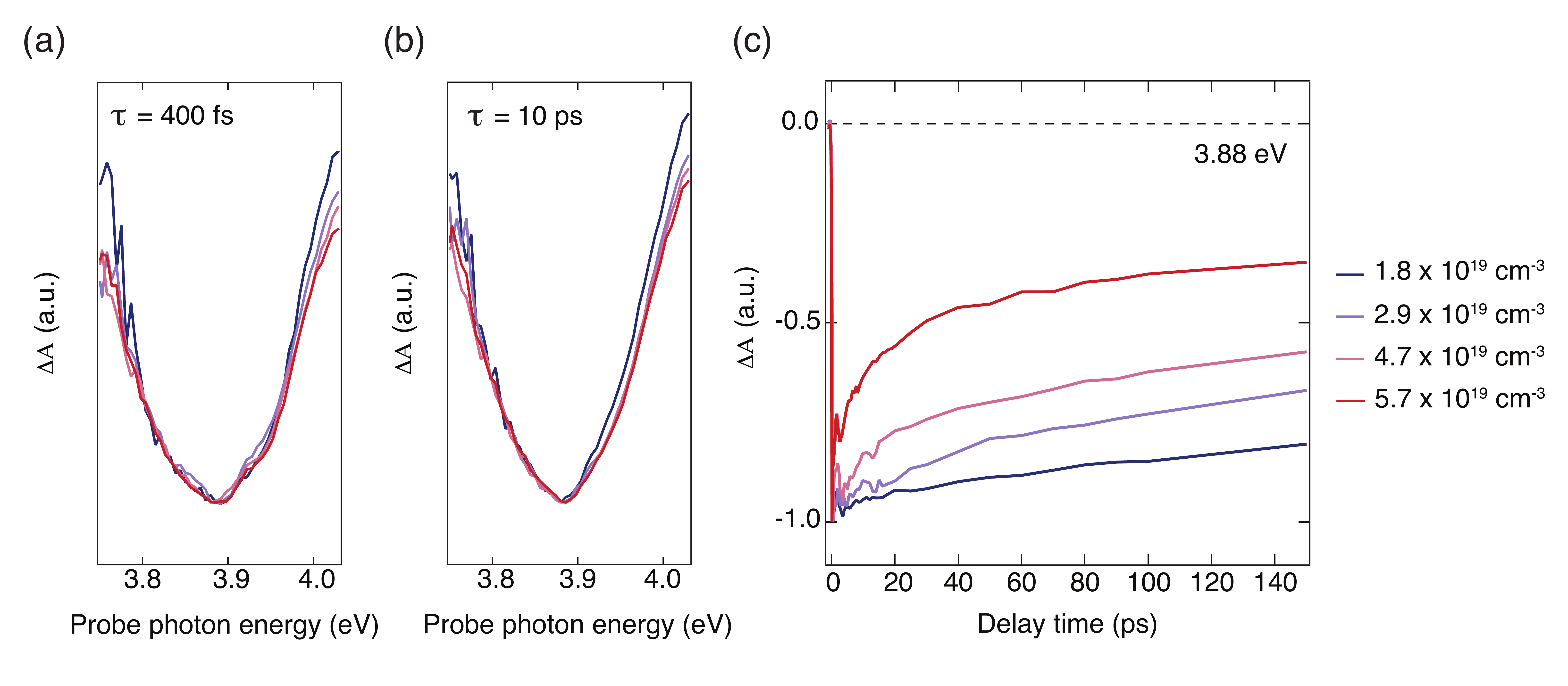}
	\caption{\textbf{(a,b)} Normalized $\Delta$A spectra as a function of probe photon energy for different photoexcited carrier densities. Spectra in \textbf{(a)} are cut at a delay time of 400 fs, spectra in \textbf{(b)} at 10 ps. \textbf{(c)} Normalized temporal traces at 3.88 eV for different photoexcited carrier densities.}
\end{figure}
\newpage

\begin{figure}[tb]
	\begin{center}
		\centering
		\includegraphics[width=\columnwidth]{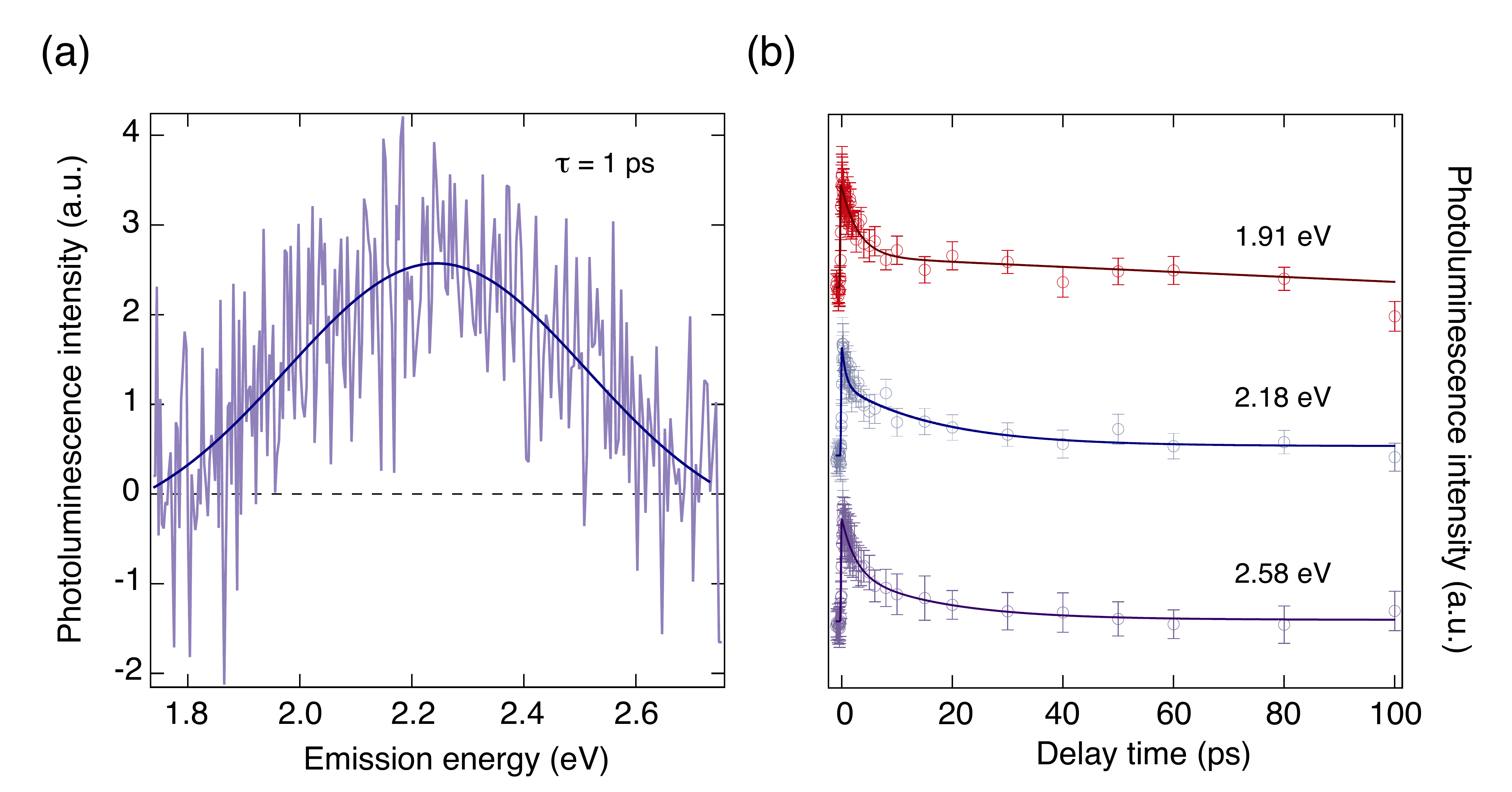}
		\caption{(a) Experimental spectrum of the ultrafast PL at a time delay of 1 ps (violet curve) and gaussian fit of the response (blue curve). (b) Temporal evolution of the ultrafast PL at 1.91 eV (red dots), 2.18 eV (blue dots) and 2.58 eV (violet dots). The solid lines are fit to the experimental curves.}
		\label{fig:PL}
	\end{center}
\end{figure}
\newpage

\begin{figure}[t]
	\centering
	\includegraphics[width=\linewidth]{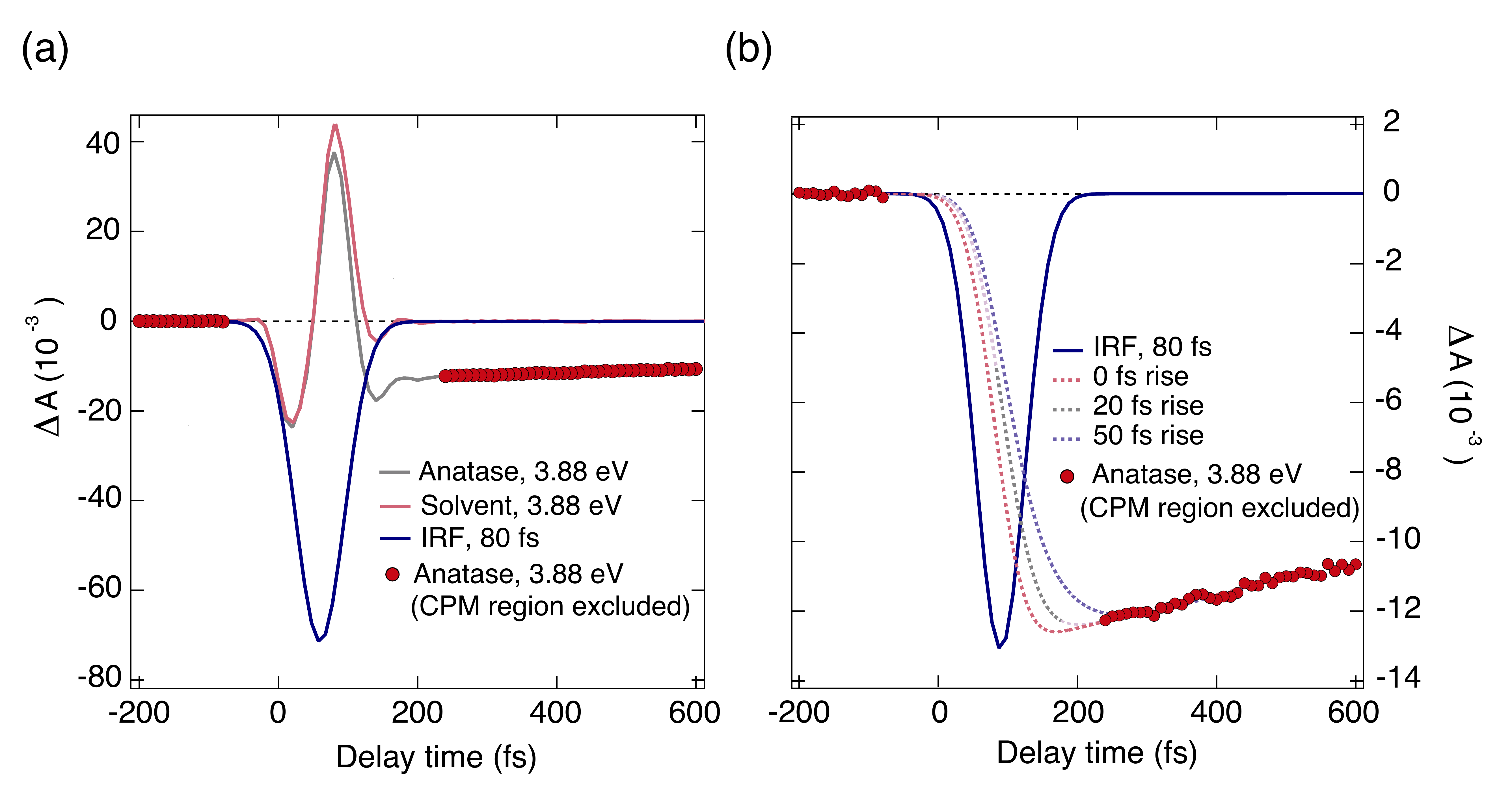}
	\caption{\textbf{(a)} Comparison of the kinetic trace of the $\Delta$A signal at 3.88 eV from the colloidal solution of anatase TiO$_2$ (grey trace, red dots) with that of the solvent alone (pink trace). The sample response with the CPM region excluded is represented by the red dots. The Gaussian IRF of 80 fs is also shown. \textbf{(b)} Evaluation of the rise-time of the $\Delta$A signal by interpolating the $\Delta$A response of anatase TiO$_2$ in the region where CPM is excluded, using a rising function (convoluted with a Gaussian IRF of 80 fs) at three different times.}
\end{figure}
\newpage

\clearpage
\newpage

\setcounter{section}{0}
\setcounter{figure}{0}
\renewcommand{\thesection}{S\arabic{section}}  
\renewcommand{\thetable}{S\arabic{table}}  
\renewcommand{\thefigure}{S\arabic{figure}} 
\renewcommand\Im{\operatorname{\mathfrak{Im}}}
\titleformat{\section}[block]{\bfseries}{\thesection.}{1em}{} 

\section{S1. Sample preparation}

The TiO$_2$ nanoparticles (NPs) were prepared using the sol-gel method~\cite{ref:mahshid}. The synthesis was carried out in a glove box under argon atmosphere. Titanium isopropoxide (Sigma Aldrich, 99.999$\%$ purity) was used as precursor and was mixed with 10 ml of 2-propanol. This mixture was added dropwise under vigorous stirring to cold acidic water ($2\,^{\circ}\mathrm{C}$, 250 ml H$_2$O, 18 M$\Omega$, mixed with 80 ml glacial acetic acid, final pH 2). At the beginning the mixture looked turbid, but after stirring it in an ice bath for 12 hours, it became transparent as the amorphous NPs were formed. The mixture was then peptized at $80\,^{\circ}\mathrm{C}$ for about 2 hours until the liquid turned into a transparent gel. The gel was autoclaved at $230\,^{\circ}\mathrm{C}$ for 12 hours. During this process the previous amorphous TiO$_2$ sample became denser and underwent a phase transition, resulting in anatase TiO$_2$ NPs. After the autoclave, the NPs have precipitated to the bottom of the container. They were separated from the supernatant and added to 100 ml acidic water (pH 2) to obtain a white colloidal solution with a final concentration of ca. 5 g/L. In Ref. \cite{rittmann2014mapping}, we reported the details of the sample characterization by means of x-ray diffraction and Transmission Electron Microscopy. Using these techniques, the good quality of the anatase phase and the spherical shape (with an average diameter of approximately 25 nm) of the NPs were demonstrated. The steady-state absorption specta reported in Fig. S1 were recorded at room temperature using a commercial spectrometer (Shimadzu, UV-3600). Before measuring the absorption spectrum of the sample, a reference spectrum of the pure solvent (acidic water, pH 2) was taken to check its transparency in the investigated spectral range.

\section{S2. Ultrafast two-dimensional deep-ultraviolet spectroscopy}

The ultrafast two-dimensional (2D) deep-ultraviolet (UV) spectroscopy measurements were performed using a set-up described in detail in Ref. \cite{ref:aubock}. Briefly, a 20 kHz Ti:Sapphire regenerative amplifier (KMLabs, Halcyon + Wyvern500), providing pulses at 1.55 eV, with typically 0.6 mJ energy and around 50 fs duration, pumps a non-collinear optical parametric amplifier (NOPA) (TOPAS white - Light Conversion) to generate sub-90 fs visible pulses (1.77 - 2.30 eV range). The typical output energy per pulse is 13 $\upmu$J. Around 60$\%$ of the output of the NOPA is used to generate the narrowband pump pulses. The visible beam, after passing through a chopper, operating at 10 kHz and phase-locked to the laser system, can be focused onto a 2 mm thick barium borate (BBO) crystal for obtaining the UV pump pulse. In this case, the pump photon energy is controlled by the rotation of the crystal around the ordinary axis and can be tuned in a spectral range up to $\geq$ 0.9 eV ($\geq$ 60 nm) wide. The typical pump bandwidth is 0.02 eV (1.5 nm) and the maximum excitation energy is about 120 nJ. The pump power is recorded on a shot-to-shot basis by a calibrated photodiode for each pump photon energy, allowing for the normalization of the data for the pump power. The remaining NOPA output is used to generate the broadband UV probe pulses with $\geq$ 100 nm bandwidth through an achromatic doubling scheme. Pump and probe pulses, which have the same polarization, are focused onto the sample, where they are spatially and temporally overlapped. The typical spot size of the pump and the probe are 100 $\times$ 150 $\upmu$m$^2$ and 40 $\times$ 44 $\upmu$m$^2$ full-width half-maximum (FWHM) respectively, resulting in a homogeneous illumination of the probed region.

For most measurements, the colloidal solution circulated into a 0.2 mm thick quartz flow-cell to prevent photo-damage and its concentration was adjusted to provide an optical density of approximately 0.4 at the pump photon energy of 4.05 eV. Alternatively, for limiting the issues of cross-phase modulation in the sample and achieving a better time resolution, the NPs solution could be made flowing in the form of 0.2 mm thick jet. The probe was measured after its transmission through the sample and its detection synchronized with the laser repetition rate. The difference of the probe absorption with and without the pump pulse has been measured at varying time delays between the pump and the probe, thanks to a motorized delay line in the probe path. After the sample, the transmitted broadband probe beam was focused in a multi-mode optical fiber (100 $\upmu$m), coupled to the entrance slit of a 0.25 m imaging spectrograph (Chromex 250is). The beam was dispersed by a 150 gr/mm holographic grating and imaged onto a multichannel detector consisting of a 512 pixel CMOS linear sensor (Hamamatsu S11105, 12.5 $\times$ 250 $\upmu$m pixel size) with up to 50 MHz pixel readout, so the maximum read-out rate per spectrum (almost 100 kHz) allowed us to perform shot to-shot detection easily. The experimental setup offered a time resolution of 150 fs, but this could be improved to 80 fs with the adoption of a prism compressor on the pump pulses, at the expenses of a reduction of the probe bandwidth in the achromatic frequency doubling scheme.

To study the response of the anatase TiO$_2$ NPs at high fluence upon excitation at 3.10 eV, we also used another visible pump/deep-UV continuum probe setup, described in Ref. \cite{consani2009vibrational}. Briefly, a 1 kHz regenerative amplifier provides 30 fs pulses at 1.55 eV with a pulse energy of 620 $\upmu$J. Part of the beam is focused onto a 2 mm thick BBO crystal for obtaining the 3.10 eV pump pulse. The remaining part of the regenerative amplifier output is used to pump one home-built NOPA, which deliver typical pulses of 1-2 $\upmu$J and 80 nm bandwidth at 600 nm, with a pulse duration of about 30 fs. The output of the NOPA is used to generate a broadband UV pulse via achromatic doubling in a 0.15 mm thick BBO crystal. The experimental scheme here is quite different from the one described above, and the bandwidth of the generated broadband pulses is narrower. Three separate experiments are necessary to fully cover the 3.20-4.60 eV spectral region. After the sample, the probe beam is dispersed by an 830 gr/mm transmission grating and focused on a 512 pixel photodiode array. Typical pump and probe focus dimensions are 120 $\upmu$m and 60-80 $\upmu$m.

\section{S3. Ultrafast broadband fluorescence upconversion}

The ultrafast broadband fluorescence upconversion measurements were performed using a set-up described in detail in Ref. \cite{cannizzo}. In our experimental configuration, a 150 kHz Ti:Sapphire regenerative amplifier system (Coherent, RegA-9000) is seeded by a mode-locked Ti:Sapphire oscillator (Coherent, Mira-SEED) and produces 4 $\mu$J pulses at 1.55 eV with a 80 fs pulse width. The beam is then split into two arms with a pellicle beam splitter: The reflected part is used for gating, the remaining for generating the pump beam at 4.66 eV via frequency tripling. The excitation light is focused with an area of 35 $\mu$m of diameter onto a 0.2 mm thick quartz flow-cell, in which the colloidal solution of anatase TiO$_2$ circulates continuously. The light emitted by the sample is collected by a parabolic mirror in forward-scattering geometry, and directed to a second mirror that focuses it onto a BBO crystal for sum frequency generation. The luminescence is then up-converted by mixing with the gate pulse in a slightly non-collinear geometry. The up-converted signal is spatially filtered and detected with a spectrograph and a liquid nitrogen cooled charge-coupled device camera. The experimental setup offers a time resolution of 150 fs.

\section{S4. Estimation of the carrier density}

We calculate the number of excited electron-hole pairs per unit cell of anatase TiO$_2$ by taking into account a sample of NPs, flowing in a quartz flow cell of thickness $l$ = 0.2 mm with a concentration of $c$ = 0.106 g/L and an optical density of $OD$ = 0.4. Specifically, we consider the ultrafast broadband deep-UV experiment in which the sample is pumped with a photon energy $E_{pump}$ = 4.05 eV at a repetition rate of $f$ = 20 kHz and average power $P$. This calculation can be easily extended for the other experimental parameters used in this work.

The energy per pulse is $E_{pulse} = P/f$, corresponding to a number of photons per pulse of $N_{pulse} = E_{pulse}/E_{ph}$. The pump is focused on an area of $A_{foc}$ = $\pi$ $\times$ 50 $\upmu$m $\times$ 75 $\upmu$m = 1.18 $\times$ 10$^{-8}$ m$^2$, corresponding to a focusing volume of $V_{pump} = A_{foc} l = 2.36 \times 10^{-12}$ m$^3$. Since the radius of one NP is $R$ $\sim$ 14 nm, the volume can be estimated by approximated the NP to a sphere. This yields $V_{np} = \frac{4}{3}\pi R^3 = 1.15 \times 10^{-23}$ m$^3$. The anatase TiO$_2$ density is $\rho$ = 3.9 $10^6$ g/m$^3$ and the mass of one NP is $M_{np} = V_{np} \rho = 4.5 \times  10^{-17}$ g. Given the concentration $c$, the number of NPs in the volume where the pump beam is focused is $N_{foc} = c V_{pump} 10^3/ M_{np} = 5.57 \times 10^6$. The number of photons absorbed in a pulse is $N_{abs}=(1-T) N_{pulse}$, where $T = 10^{-OD}$ = 0.398. In conclusion, the photoexcited carrier density $n$ is calculated as the ratio between total number of absorbed photons and the total illuminated volume $N_{foc} \times V_{pump}$.

\section{S5. Indirect bandgap photoexcitation}

We also perform a separate ultrafast broadband deep-UV experiment to verify the appearance of the a-axis exciton feature when the pump beam is tuned within the spectral region of indirect interband transitions. These transitions involve the phonon-assisted generation of electron-hole pairs in the material. While the electrons are promoted in the region around the $\Gamma$ point of the conduction band, the holes are expected to be located in the valence band at the X and Z valleys of the Brillouin zone. As such, the cross section of these transitions is rather low. Spectrally, the region of phonon-assisted interband transitions extends from the indirect bandgap energy 3.20 eV to the direct exciton feature and, as such, it lies below the excitation region explored in the 2D deep-UV measurements. To this aim, in this separate experiment, we tune our pump photon energy to 3.54 eV and measure the $\Delta$A response of the anatase TiO$_2$ colloidal solution in the vicinity of the a-axis exciton peak (3.70 - 4.10 eV). The photoexcited carrier density is estimated around 1 $\times$ 10$^{19}$ cm$^{-3}$, which yield a very weak ($\Delta$A $\sim$ 10$^{-3}$) signal. The colour-coded map of $\Delta$A as a function of probe photon energy and time delay between pump and probe is shown in Fig. S2. We clearly distinguish the a-axis exciton bleach persisting up to 1 ns.

\section{S6. Choice of the electron effective mass}

The electron effective mass was recently measured via ARPES to be around $m_e^*$ = 0.7$m_e$ (where $m_e$ is the bare electron mass) for excess carrier densities $n$ $\sim$ 10$^{18}\div$10$^{19}$ cm$^{-3}$ \cite{ref:moser}, which is larger than the one predicted by band structure calculations ($m_e^*$ = 0.42$m_e$ \cite{hitosugi2008electronic}) due to the polaronic dressing of the electrons. For higher carrier density regimes, $m_e^*$ is expected to decrease towards the values predicted by band structure calculations due to an effective screening of the electron-phonon interaction \cite{ref:moser}. Thus, in Eqs. (2)-(3) of the main text, we make use of the value reported by band theory.

\section{S7. Ultrafast photoluminescence}

To establish the nature of the mechanisms governing the recombination dynamics in anatase TiO$_2$ NPs, we perform ultrafast broadband fluorescence upconversion. Our aim is to determine whether radiative (bimolecular) recombination processes take place at room temperature within the first 100 ps and, in this case, to evaluate the effective photoluminescence (PL) quantum yield for such processes. In the past, a number of works studied the radiative decay in anatase TiO$_2$ single crystals, thin films and NPs, both at low and room temperature \cite{ref:tang_PL, watanabe2000time, watanabe2005time, wakabayashi2005time, harada2007time, cavigli2009volume, cavigli2010carrier, preclikova2010nanocrystalline}. Since anatase TiO$_2$ is an indirect bandgap insulator, band-to-band radiative recombination processes are unlikely, as the photoluminescence process requires phonon emission to conserve momentum. As such, the PL response in anatase TiO$_2$ is dominated by other sources of radiative recombination. More specifically: i) the intrinsic PL from the radiative recombination of self-trapped excitons; ii) the extrinsic PL from the radiative decay at impurities and defects. As both contributions give rise to broad PL signals covering the visible range, a distinction can be made depending on the central PL photon energy. Intrinsic PL from self-trapped excitons is prominent at low temperature and shows a large Stokes-shift with respect to the indirect absorption threshold of the material. This PL band is centered around 2.37 eV at low temperature and undergoes a sizeable blueshift for increasing temperature, reaching 2.57 eV at 300 K. In contrast, broad PL signals from defects show less pronounced temperature dependence for their central peak, being centered around 2.24 eV. Interestingly, no studies of the ultrafast PL in the first 100 ps of the response have been reported yet. Hence, it is still an open question whether the system undergoes radiative recombination pathways during these short timescales. An estimate of the PL quantum yield would also clarify the origin of the high order recombination processes observed in the ultrafast 2D deep-UV data. 

To this aim, we performed ultrafast broadband fluorescence upconversion measure- ments on the same colloidal solution of anatase TiO$_2$ NPs used for the ultrafast 2D deep-UV spectroscopy study. A detailed study of the origin of this ultrafast PL response will be subject of a future work. Here we are only interested in ruling out any involvement of radiative decay channels in the recovery of the exciton bleach signal shown in the main text. Thus, irrespective of the origin (intrisic vs. extrinsic) of the PL band, we evaluate the PL quantum yield for time delays below 500 ps. By considering that the whole PL band observed in our experiment yields 3.6 $\times$ 10$^9$ emitted photons per second. This value has to be compared with the number of absorbed photons per second by the whole colloidal anatase TiO$_2$ solution, which amounts to 2.8 $\times$ 10$^{15}$. Thus, we estimate a quantum yield of $\sim$ 1.3 $\times$ 10$^{-6}$. This is consistent with the fact that the quantum yield of anatase TiO$_2$ NPs at room temperature and under steady-state conditions has been found $\sim$ 2.5 $\times$ 10$^{-3}$ \cite{abazovic2006, abazovic2008}. Indeed, the latter represents an upper limit to the PL quantum yield that can be obtained in anatase TiO$_2$ NPs at room temperature and, as expected, it implies that the radiative decay channel is not efficient in this material.
\newpage

\clearpage
\newpage

\begin{figure}[tb]
	\begin{center}
		\centering
		\includegraphics[width=0.5\columnwidth]{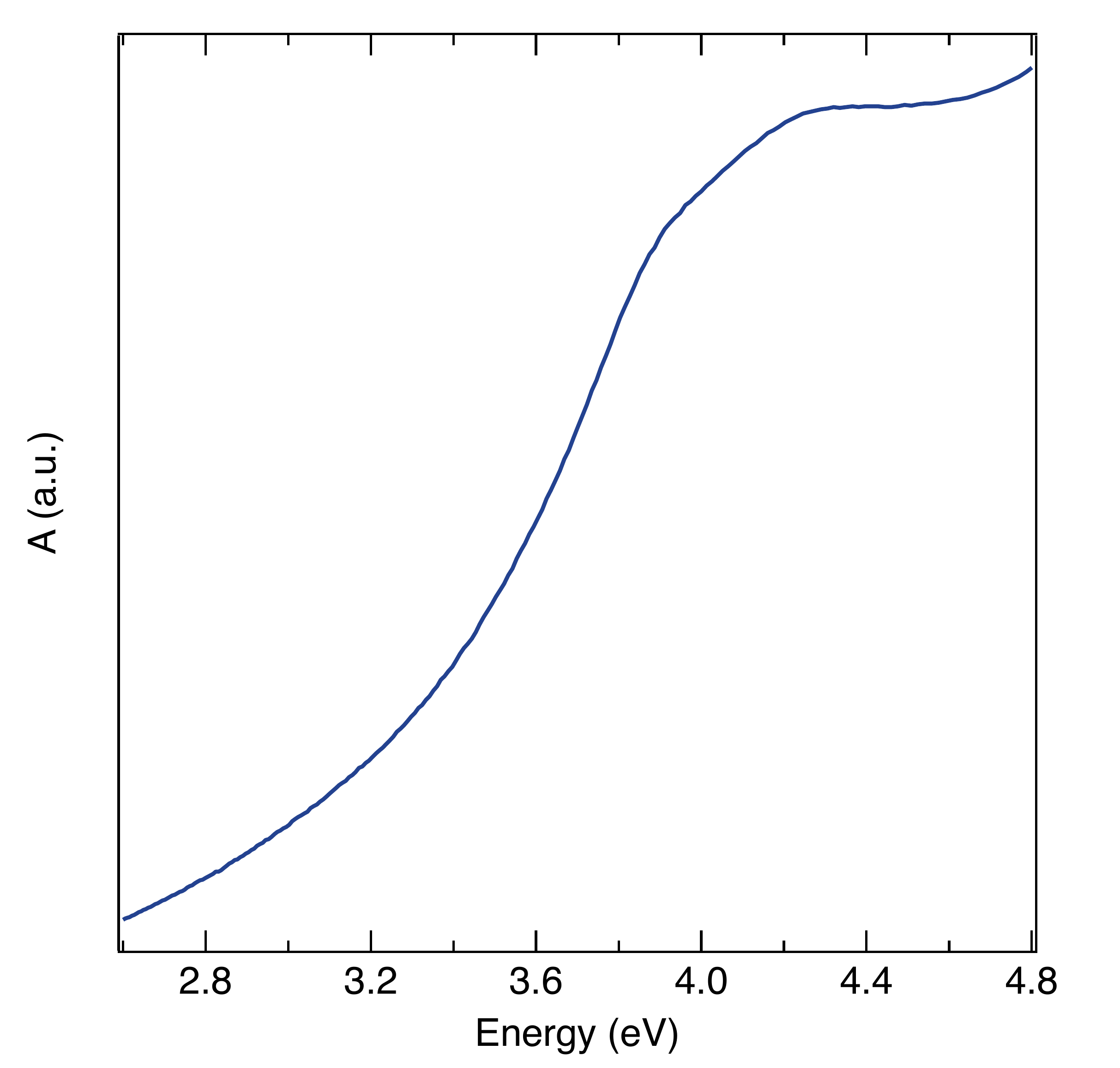}
		\caption{Room temperature steady-state absorption spectrum of anatase TiO$_2$ NPs dispersed in aqueous solution.}
		\label{fig:350nm}
	\end{center}
\end{figure}
\newpage
\clearpage

\begin{figure}[tb]
	\begin{center}
		\centering
		\includegraphics[width=0.5\columnwidth]{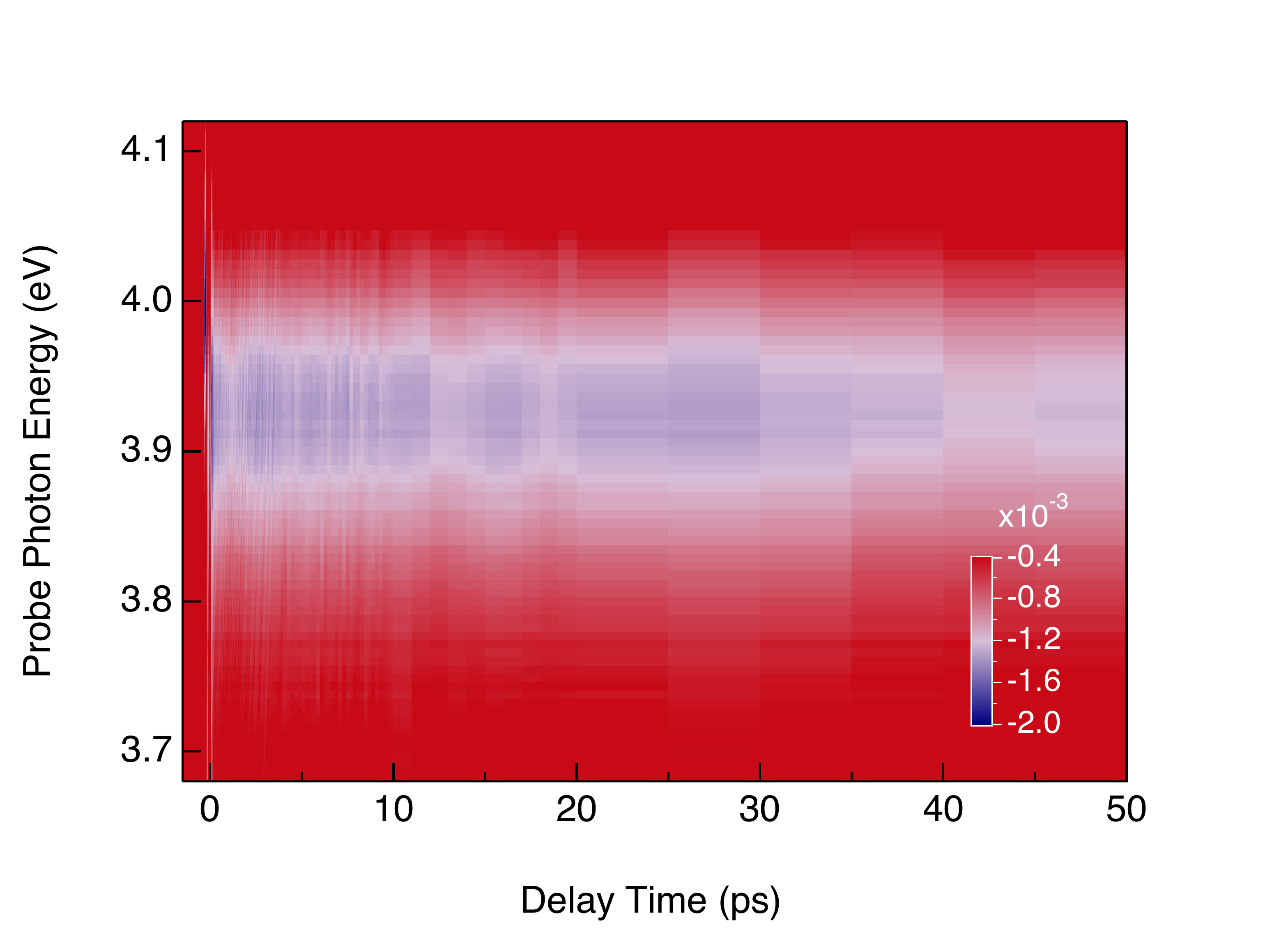}
		\caption{Colour-coded map of $\Delta$A measured on a colloidal solution of anatase TiO$_2$ NPs at room temperature as a function of probe photon energy and time delay between pump and probe. The time resolution is estimated 150 fs, the pump photon energy is set at 3.54 eV and the delivered pump fluence is 100 $\mathrm{\mu J/cm^2}$.}
		\label{fig:350nm}
	\end{center}
\end{figure}
\newpage
\clearpage

\begin{figure}[t]
	\centering
	\includegraphics[width=0.5\linewidth]{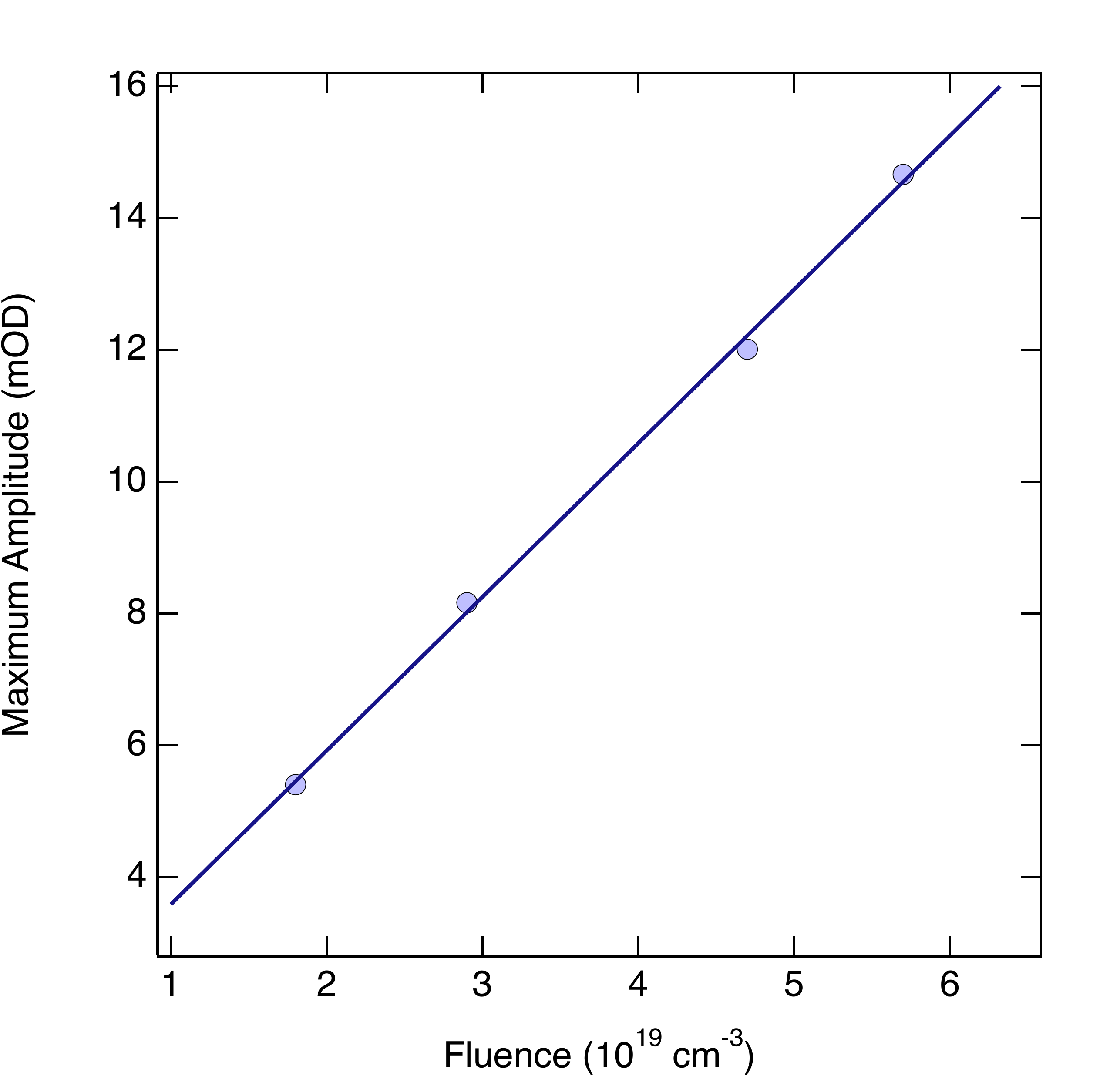}
	\caption{Maximum intensity of the $\Delta$A signal as a function of the absorbed fluence.}
\end{figure}
\newpage

\clearpage

\providecommand{\noopsort}[1]{}\providecommand{\singleletter}[1]{#1}%

\end{document}